\documentclass{aa} 
\usepackage{natbib,amssymb,graphicx}
\bibpunct{(}{)}{;}{a}{}{,} 

\begin{document}

\title{Dust and the spectral energy distribution of the OH/IR star OH
  127.8+0.0: Evidence for circumstellar metallic iron\thanks{Based on
    observations with ISO, an ESA project with instruments funded by
    ESA Member States (especially the PI countries: France, Germany,
    the Netherlands and the United Kingdom) and with the participation
    of ISAS and NASA.}}  
\subtitle{} 
\titlerunning{The spectral energy
  distribution of OH 127.8+0.0}

\author{F. Kemper\inst{1} \and A. {de Koter}\inst{1} \and L.B.F.M.
  Waters\inst{1,2} \and J. Bouwman\inst{1}\thanks{\emph{present
      address:} Service d'Astrophysique, CEA/DSM/DAPNIA, C.E.~Saclay,
    F-91191 Gif-sur-Yvette, France} \and A. G. G. M.
  Tielens\inst{3,4}}

\institute{Astronomical Institute ``Anton Pannekoek'', University of
  Amsterdam, Kruislaan 403, 1098 SJ Amsterdam, The Netherlands \and
  Instituut voor Sterrenkunde, Katholieke Universiteit Leuven,
  Celestijnenlaan 200B, B-3001 Heverlee, Belgium \and Kapteijn
  Institute, University of Groningen, P.O. Box 800, 9700 AV Groningen,
  The Netherlands \and SRON Laboratory for Space Research, P.O. Box
  800, 9700 AV Groningen, The Netherlands}

\offprints{F. Kemper, (ciska@science.uva.nl)}

\date{Received / Accepted}

\abstract{We present a fit to the spectral energy distribution of
  \object{OH 127.8+0.0}, a typical asymptotic giant branch star with
  an optically thick circumstellar dust shell.  The fit to the dust
  spectrum is achieved using non-spherical grains consisting of
  metallic iron, amorphous and crystalline silicates and water ice.
  Previous similar attempts have not resulted in a satisfactory fit to
  the observed spectral energy distributions, mainly because of an
  apparent lack of opacity in the 3--8 $\mu$m region of the spectrum.
  Non-spherical metallic iron grains provide an identification for the
  missing source of opacity in the near-infrared.  Using the derived
  dust composition, we have calculated spectra for a range of
  mass-loss rates in order to perform a consistency check by
  comparison with other evolved stars. The $L-[12$ $\mu$m$]$ colours
  of these models correctly predict the mass-loss rate of a sample of
  AGB stars, strengthening our conclusion that the metallic iron
  grains dominate the near-infrared flux. We discuss a formation
  mechanism for non-spherical metallic iron grains.   }

\maketitle

\section{Introduction}
\label{sec:introduction}

Oxygen-rich Asymptotic Giant Branch (AGB) stars show an infrared
excess on their spectral energy distribution (SED), which arises from
thermal emission of dust located in a circumstellar shell. It is well
established that the dust in this shell mainly consists of silicates,
deduced from the clear detection of the 10 and 18 $\mu$m features.
These bands are due to the Si-O bond stretching and Si-O-Si bond
bending vibrations, respectively.  It was noticed already some 25
years ago \citep{JM_76_dust,B_77_dustshells} that the opacities of
various types of silicates measured in the laboratory, were not high
enough to explain the shape of the spectrum in the near-infrared (NIR,
$3 < \lambda < 8$ $\mu$m). \citet{JM_76_dust} investigated the effect
of pure silicates with graphite inclusions and of meteoritic rock on
the shape of the SED. These materials were taken to represent
silicates with various metallic inclusions, and referred to as
\emph{dirty silicates}. Radiative transfer calculations using the
opacity of dirty silicates were in good agreement with the
observations.  \citet{DL_84_optprop} used laboratory measurements of
limited resolution and wavelength coverage \citep{HS_73_optpropsolids}
in which the optical constants are partly modified and expanded in
wavelength regime to match the astronomical observations of
interstellar and circumstellar silicates
\citep{JM_76_dust,RMC_83_mucep} These optical constants are usually
referred to as \emph{astronomical silicate}.  The exact chemical
composition of this astronomical silicate is not known, nevertheless
it is widely used in theoretical studies of the radiative transfer in
circumstellar dust shells
\citep[e.g.][]{B_87_dustshells,ST_89_theory,JT_92_SED,LL_93_SED}.
Recently there have been efforts to derive improved optical constants
for astronomical silicates, both for lines-of-sight in the
interstellar medium (ISM) and toward oxygen-rich evolved stars
exclusively \citep{OHM_92_cosmicsilicates,DP_95_cssil,S_99_optprop}.
These optical constants still do not provide the answer to the
composition and the nature of circumstellar silicates.

The study of circumstellar dust shells in the pre-ISO era was based on
low resolution spectroscopy, but with the launch of the Infrared Space
Observatory (ISO) \citep{KSA_96_ISO}, a large wavelength region, from
2--200 $\mu$m, became available for intermediate resolution
spectroscopic observations ($\lambda / \Delta \lambda \gtrsim 400$).
The ISO spectra of oxygen-rich evolved stars turned out to be
extremely rich in solid state features. Hence, it became meaningful to
attempt spectral fits using properly characterized cosmic dust
analogs, rather than astronomical silicates. \citet{DDW_00_OHIR} aim
to do this for OH/IR stars (i.e.~AGB stars with an optically thick
dust shell), but their work is not complete and parts of the spectrum
are not fitted to a satisfactory level.  \citet{HMD_01_VYCMa} have
studied the optically thin dust shell surrounding red supergiant
\object{VY CMa} and are able to fit the SED, albeit after including a
large overabundance of metallic iron in the dust composition.

In this study, we aim to determine the dust composition with full
radiative transfer calculations using exclusively optical constants of
well-defined materials measured in the laboratory, examining all
astronomically relevant materials, and taking abundance constraints
into account. Revealing the exact dust composition is an important
step toward understanding dust formation and processing in the
outflows of oxygen-rich evolved stars.

The paper is organized as follows: In Sect.~\ref{sec:near-infr-probl}
we discuss the spectral energy distribution arising from full
radiative transfer calculations using amorphous olivine as the only
dust component, and compare it to the observations.
Sect.~\ref{sec:improvedfit} discusses the dust composition and grain
properties required to improve the fit to the observations. A spectral
fit to \object{OH 127.8+0.0} is presented in Sect.~\ref{sec:oh127},
supported by a consistency check on other AGB stars in
Sect.~\ref{sec:lmin12}.  Section~\ref{sec:disc-concl} contains a
discussion of the results.

\section{Modelling the circumstellar environment}
\label{sec:near-infr-probl}

In this section we describe and discuss the results from the radiative
transfer calculations assuming that the dust only consists of
amorphous olivine. We describe the discrepancies between the
calculated and observed spectral energy distributions of AGB stars.

\subsection{Model assumptions and default grid}
\label{sec:basics}

Our approach is to calculate a set of model spectra of dust shells
characteristic for both Mira and OH/IR stars, and to compare these
with ISO observations. The stellar and wind parameters are the same
for all models in the grid, except for the mass-loss rate which is
varied from relatively low values, typical for Mira's, to relatively
large values, characteristic for OH/IR stars. This leads to a variety
of optical depths, which can be compared to the optical depth in the
10 $\mu$m feature of observed spectra. A similar approach has been
taken by \citet{B_87_dustshells} who also studied the appearance of
the SED of AGB stars as a function of mass loss, comparing the results
with IRAS observations and ground based infrared data.

We use the code {\sc modust} to model the spectrum of the
circumstellar dust shell. The radiative transfer technique applied in
this code has been outlined by \citet{KWD_01_xsilvsmdot}. The
specification of grain properties, such as size and shape
distribution, are discussed in \citet{BDV_00_ABAur}. Here, we only
discuss details relevant for the presented models.  The dust is
assumed to be distributed in a spherical shell, with inner radius
$R_{\mathrm{in}}$ and outer radius $R_{\mathrm{out}}$, which is
irradiated by a central star with $T_{\mathrm{eff}} = 2.7 \cdot 10^3$
K, a radius $R_{\star} = 372\, R_{\sun}$ and a luminosity $L_{\star} =
6.3 \cdot 10^{3}\, L_{\odot}$, thus describing a typical AGB star.
The input spectrum of the central star is not simply a black body but
is a characteristic mean spectrum for spectral type M9III, inferred
from observations (380 nm $\lesssim \lambda \lesssim$ 900 nm) and
extended with synthetic spectra in the range 99 -- 12500 nm
\citep{FPT_94_Mstars}. The M9III star with a $T_{\mathrm{eff}}=3126$
K, emits significantly less flux at $\lambda < 1$ $\mu$m than a black
body of the same temperature. Dust is efficiently heated in this
wavelength region, and therefore the dust located at the inner edge of
the dust shell around an M9III star would be somewhat cooler than the
dust at the inner radius around a black body of 3126 K.  This
temperature difference decreases with distance from the central star.
In case of an optically thin dust shell, the observed spectrum is
dominated by the thermal emission of the warm dust at the inner
radius. Therefore, the silicate features are somewhat stronger in case
the black body is used as a central source. In case of an optically
thick dust shell, there is no difference in the spectral energy
distribution between the two possibilities, because only the dust in
the outer regions is visible for the observer.

The adopted mass-loss rates range between 5 $\cdot$ 10$^{-8}$ and 1
$\cdot$ 10$^{-3}$ $M_{\odot}$ yr$^{-1}$. The inner radius of the dust
shell is defined by the condensation temperature of amorphous
silicate, taken to be $T_{\mathrm{cond}} = 900$~K
\citep{B_87_dustshells}.  The exact location of the inner radius is
determined by the largest grains and is a function of the mass-loss
rate as the dust density in the shell determines the diffuse component
of the radiation field, also known as backwarming, thus constraining
the temperature profile.  The values for $R_{\mathrm{in}}$ range from
6.05 to 29 $R_{\star}$ for the smallest and largest mass loss,
respectively (see Tab.~\ref{tab:innerradius}).  The outer radius is
chosen to scale correspondingly to 100 $R_{\mathrm{in}}$.  This is
significantly lower than what is used by others
\citep[e.g.][]{B_87_dustshells}, but higher values lead to an
overestimation of the flux around 20 -- 25 $\mu$m.  The dust density
in the shell is determined using a constant outflow velocity of 20 km
s$^{-1}$, implying the density decreases as $\rho(r) \propto r^{-2}$.
We used a dust/gas mass ratio of 0.01.

\begin{flushleft}
\begin{table}
\caption[]{Inner radius of the dust shell used in our initial model 
calculations, for different mass-loss rates, based on the work
of \citet{B_87_dustshells}.}
\begin{tabular}{l l }
\hline
\hline
$\dot{M}$ ($M_{\odot}$ yr$^{-1}$) & $R_{\mathrm{in}}$ ($R_{\ast}$)\\
\hline
5($-$8) & 6.05 \\
1($-$7) & 6.20 \\
2($-$7) & 6.37 \\
5($-$7) & 6.56 \\
1($-$6) & 6.91 \\
2($-$6) & 7.26 \\
5($-$6) & 8.23 \\
1($-$5) & 9.17 \\
2($-$5) & 10.2 \\
5($-$5) & 12.3 \\
1($-$4) & 14.3 \\
2($-$4) & 17.8 \\
5($-$4) & 23.5 \\
1($-$3) & 29.0 \\
\hline
\hline
\end{tabular}
\label{tab:innerradius}
\end{table}
\end{flushleft}

To obtain the absorption and scattering efficiencies we used optical
constants of an amorphous olivine (MgFeSiO$_4$) sample measured by
\citet{DBH_95_glasses} and assumed that the grains are spherical, such
that Mie theory can be applied
\citep[see][]{V_57_lightscattering,BH_83_scattering}.  The size
distribution of grains was assumed to be interstellar, i.e. given by
$n(a) \propto a^{-3.5}$, where $a$ is the grain size
\citep{MRN_77_grainsize}.  For the circumstellar silicates we adopted
a minimum and maximum size of the grains of 0.01 and 1 $\mu$m.

\subsection{Problems in fitting the SED of AGB stars}
\label{sec:problems}

\begin{figure}
  \resizebox{\hsize}{!}{\includegraphics{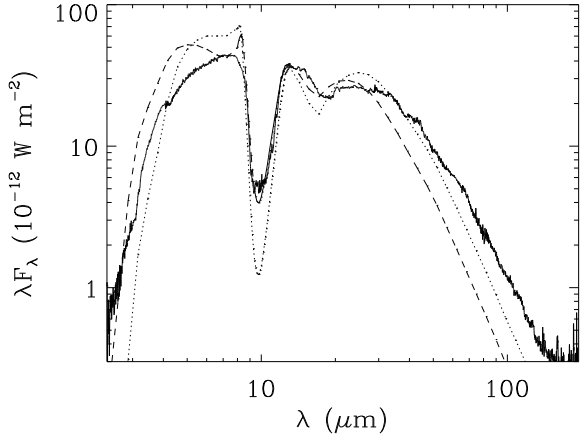}}
\caption{ISO spectrum of \object{OH 127.8+0.0} (solid line), 
  together with two model spectra. The dashed line is the SED of the
  model with $\dot{M} = 1 \cdot 10^{-4}$ $M_{\odot}$ yr$^{-1}$, and
  the dotted line corresponds to the model with $\dot{M} = 2 \cdot
  10^{-4}$ $M_{\odot}$ yr$^{-1}$.  Note the discrepancies between the
  model fits and the observed spectrum at $\lambda =$ 3--8 $\mu$m
  region, at the peak position of the 18 $\mu$m feature and at the
  long wavelength part of the spectrum.}
\label{fig:NIRproblem}
\end{figure}

Figure~\ref{fig:NIRproblem} shows a comparison of two of our predicted
SEDs with the ISO spectrum of the oxygen-rich star \object{OH
  127.8+0.0} (solid line). The models with mass loss $1 \cdot 10^{-4}$
(dashed line) and $2 \cdot 10^{-4}$ $M_{\odot}$ yr$^{-1}$ (dotted
line) were selected. The first because it fits the 10 $\mu$m region
best; the second because it gives teh best overall shape of the
spectrum, even though the discrepancies remain large. The poor
agreement between the model spectrum and observations illustrates the
problem of fitting AGB spectra. The main discrepancies are: {\bf i) }
the NIR flux is overestimated.  {\bf ii) } the shape and predicted
position of the olivine resonance at 18 $\mu$m are not well
reproduced, complicating the fit to the mid-infrared (MIR, $8 <
\lambda < 30$ $\mu$m) spectrum.  Specifically, we predict the center
of this absorption feature at 16.3 $\mu$m.  {\bf iii)} The flux in the
far-infrared (FIR, $\lambda > 30$ $\mu$m) is underestimated, and the
models show a steeper spectral slope than observed.  In
Sect.~\ref{sec:improvedfit} these problems will be discussed and
resolved. All three issues can be explained in terms of dust
composition and grain properties.

\section{Improvements to the spectral fit}
\label{sec:improvedfit}

\subsection{Metallic iron as a source of NIR opacity}
\label{sec:fe}

\begin{figure}
  \resizebox{\hsize}{!}{\includegraphics{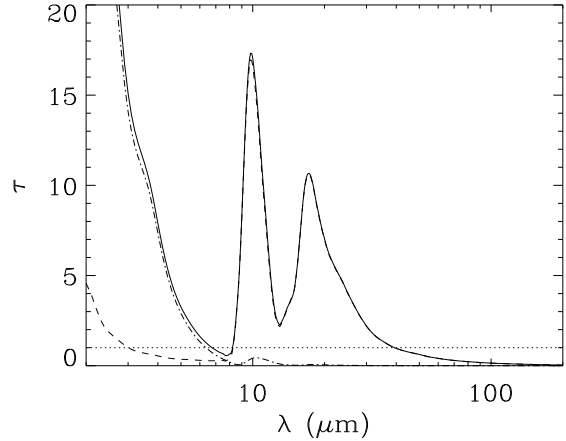}}
\caption{Optical depth towards the central star in the dust shell of an 
  evolved star with a gas mass-loss rate 1 $\cdot$ 10$^{-4}$
  $M_{\odot}$ yr$^{-1}$. The used optical constants are from amorphous
  olivine \citep{DBH_95_glasses}. $\tau = 1$ is indicated with the
  dotted line.  The dashed line indicates absorption and the
  dashed-dotted line denotes scattering. The solid line represents the
  combined optical depth due to both effects. Note that the combined
  optical depth is dominated by scattering for $\lambda < 8$ $\mu$m
  and by absorption for $\lambda > 8$ $\mu$m.}
\label{fig:tau}
\end{figure}

Let us first focus on the behaviour of the two predicted spectra. In
both models the radial monochromatic {\em absorption} optical depth in
the NIR is about equal to or somewhat larger than unity; in the 10 and
18 $\mu$m feature the dust medium is optically thick, while in the FIR
it is optically thin.  This is shown in Fig.~\ref{fig:tau}, where the
optical depth is given for a typical dust shell consisting of only
amorphous olivine.  In general, the flux at short wavelengths
decreases with increasing density as a larger fraction of the thermal
radiation emitted by the hottest grains suffers additional
thermalization in colder regions of the dust medium. So, one way to
further reduce the NIR flux is simply to increase the mass-loss rate.
However, this will also increase the optical depth at 10 and 18
$\mu$m, and by that will overpredict the absorption strength of these
features.  Thus, varying the physical parameters, such as the density
distribution, the inner radius and the outer radius, will not cause a
wavelength dependent effect on the optical depth.  In order to resolve
the NIR problem one needs to selectively increase the opacity at these
wavelengths, i.e. a species is missing of which the absorption
efficiency follows a {\em smooth} distribution -- as no unidentified
resonances are present -- predominantly contributing in the $3-8$
$\mu$m region. Using the model grid calculations to fit the optical
depth at various wavelengths, we can estimate that
$\tau_{\mathrm{NIR}} / \tau_{10\,\mu\mathrm{m}} \approx 1/5$.

\begin{figure}
  \resizebox{\hsize}{!}{\includegraphics{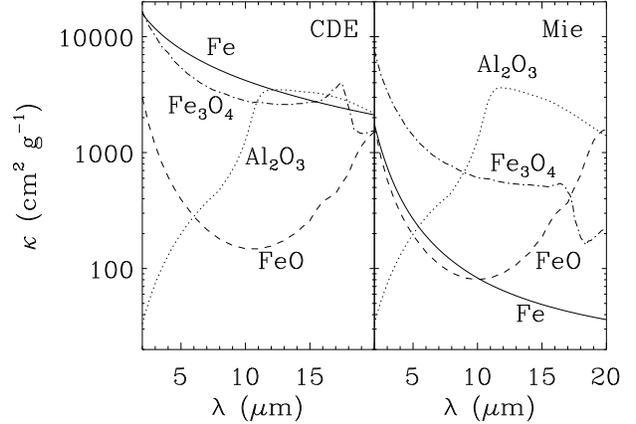}}
\caption{Mass absorption coefficients  $\kappa$ (cm$^{2}$ g$^{-1}$)
  as a function of wavelength for different species contributing in
  the NIR. The left panel shows the calculations for a continuous
  distribution of ellipsoids (CDE) which supposedly represents
  non-spherical particles, and the right panel shows the Mie
  calculations, used for spherical particles. The labels indicate the
  mass absorption coefficient curves for the considered species.  Note
  the large differences between the mass absorption coefficients for
  CDE and Mie calculations in case of metallic iron and Fe$_3$O$_4$,
  caused by their highly conductive nature.}
\label{fig:qval}
\end{figure}

For chemistry and abundance reasons we considered the following dust
species which could be present in oxygen-rich dust shells: metallic Fe
\citep{OBA_88_Fe}, FeO \citep{HBM_95_oxide}, SiO$_2$
\citep{G_63_physhandbook,D_36_quartz,SK_61_quartz,P_85_sio2,LSH_88_aerosol},
Fe$_3$O$_4$ \citep{S_74_optprop}, Al$_2$O$_3$
\citep{KKY_95_corundum,BDH_97_12-17um} and amorphous and crystalline
H$_2$O ice \citep{BLW_69_ice,W_84_ice}. The mass absorption
coefficients and extinction efficiencies have been determined from the
optical constants. The missing source of opacity has to contribute
significantly to the continuum in the 3--8 $\mu$m region, but should
not show strong and narrow features in the 2--20 $\mu$m region.  This
eliminates SiO$_2$ and water ice for this purpose.  SiO$_2$ has very
strong resonances at $\sim$8.7, $\sim$12.4, $\sim$14.4 and $\sim$20.5
$\mu$m, while the contribution to the continuum in the 3--8 $\mu$m
region is orders of magnitude less. These sharp resonances are not
observed, therefore only a small amount of SiO$_2$ can be present in
the dust shell, which is not enough to significantly increase the NIR
opacity. Water ice, both amorphous and crystalline, has features at
$\sim$3.1, $\sim$4.4, $\sim$6.3 and $\sim$12 $\mu$m.  In between these
features, the continuum is relatively weak. The $\sim$3.1 micron
feature is the strongest feature and is actually seen in absorption in
most AGB spectra. The amount of water ice required to fit this feature
explains only a negligible fraction of the NIR opacity.
Fig.~\ref{fig:qval} shows the mass absorption coefficients of the
remaining candidates, with broad band or continuum emission in the NIR
wavelength region, calculated for both spherical and non-spherical
particles. A continuous distribution of ellipsoids \citep[CDE,
see][]{BH_83_scattering} is used to represent non-spherical grains.
The strongest contribution in the 3--8 $\mu$m region is caused by
non-spherical metallic iron particles, suggesting these are a likely
candidate for the missing opacity in the NIR in the spectrum of AGB
stars.  The mass absorption coefficients of FeO and Al$_2$O$_3$ are
much lower in the NIR region, while the contribution at longer
wavelengths is more significant. This rules these species out as the
missing source of opacity.  Fe$_3$O$_4$ does contribute in the NIR
although a factor of 2 less than Fe, but shows spectral features in
the 15-20 $\mu$m region which are not observed in the spectrum.  In
this study we will therefore concentrate on Fe as the most likely
candidate for the moving NIR opacity, because it has the highest mass
absorption coefficient in the 3--8 $\mu$m region and the most
favourable spectroscopic properties.

\subsection{Non-spherical grains}
\label{sec:CDE}

The shift in the peak position of the 18 $\mu$m feature between model
calculations and observations is a second problem that occurs. This
can be solved using non-spherical amorphous olivine grains.  The
difference between the mass absorption coefficients of spherical and
non-sperical olivine grains is clearly shown in Fig.~2 of the work by
\citet{DDW_00_OHIR}.  In particular the shape of the 18 $\mu$m feature
is much affected and the peak position shifts to a somewhat longer
wavelength for non-spherical grains. However, \citet{DDW_00_OHIR} did
not use these non-spherical particles in their radiative transfer
calculations of optically thick dust shells.

A continuous distribution of ellipsoids (CDE) may better approximate
the actual variety of grain shapes that is present in the outflow of
evolved stars. The optical properties are dependent on the shape,
which is for example apparent in a shift of the peak position of
spectral features. Shape effects are most prominent for conducting
materials.  Strictly speaking, the CDE approximation is only valid in
the Rayleigh limit, where the size of the particles is at least a
factor of $\sim$ 20 smaller than the wavelength. As we are using CDE
calculations to solve problems with the 18 $\mu$m feature and our
maximum grain size used is 1.0 $\mu$m, this condition is not violated.
The result that metallic iron in CDE accounts for the missing NIR
opacity, is also consistent with the Rayleigh limit, because the
metallic iron grains are probably incorporated in the silicate grains
(see Sect.~\ref{sec:disc-concl}) and therefore have sizes
significantly smaller than 1.0 $\mu$m.  For an elaborate description
of resonances and the absorption characteristics of non-spherical
particles the reader is referred to \citet{BH_83_scattering}.

\subsection{Water ice features in the far-infrared}
\label{sec:waterice}

The improved fit to the NIR part of the SED leads to a redistribution
of the radiation such that the flux in the NIR is suppressed and FIR
flux levels are increased. Additional improvement of the FIR flux
might be achieved by adjusting the outer radius of the dust shell,
where cold dust -- optically thin in the FIR -- contributes.  The
remaining discrepancy between the observed and modelled flux levels at
the long wavelength side can probably be overcome by water ice as an
additional dust component.  The presence of the 3, 43 and 60 $\mu$m
features in most OH/IR stars implies that crystalline water ice is an
important dust component \citep{SKB_99_ohir}. The 3 $\mu$m feature is
seen in absorption (see also Sect.~\ref{sec:fe}).  In addition to the
43 and 60 $\mu$m emission features, crystalline water ice also has an
instrinsically strong underlying continuum $>30$ $\mu$m which
contributes to the FIR emission in the SED.

\section{Results for OH 127.8+0.0}
\label{sec:oh127}

\subsection{Stellar parameters and radiative transfer modelling}
\label{sec:modust}

We have constructed a model representative for AGB stars, with the
intention to fit the spectrum of \object{OH 127.8+0.0}, a high
mass-loss rate AGB star.  The input spectrum at the inner radius is
that of a star with spectral type M9III \citep{FPT_94_Mstars}, and a
radius of 372 $R_\odot$.  The density distribution and total mass of
the dust shell are determined by the velocity profile and the inner
and outer radii.  The outflow velocity was again set to 20 km
s$^{-1}$, implying a density distribution $\propto r^{-2}$. In reality
the wind probably accelerates while dust is being formed in the inner
regions, until it reaches the terminal velocity. However, the dust
shell is very optically thick in the NIR, where the warm dust in the
inner regions most effectively radiates, implying that density
variations in the inner parts -- well inside the $\tau = 1$ surface --
of the dust shell do not affect the emerging SED.  Therefore it is
impossible to further constrain the velocity profile and the inner
radius; we chose the latter to be consistent with a condensation
temperature of $\sim 900$ K, using $R_{\mathrm{in}} = 14.3 \, R_\ast$
($3.7 \cdot 10^{14}$ cm). This temperature corresponds to the
condensation temperature of amorphous silicates, but crystalline
silicates and metallic iron condense at higher temperatures
\citep{GS_99_condensation}.  We performed test calculations in which
we varied the inner radius and did not see a difference in the
resulting SED.  It is possible to determine the outer radius, found to
be at $R_{\mathrm{out}} = 5000 \, R_\ast$ ($1.3 \cdot 10^{17}$ cm),
with an accuracy of 15\%, by fitting the slope in the FIR.  By fitting
the optical depth of the 10 $\mu$m silicate feature, the dust
mass-loss rate was found to be $7 (\pm 1) \cdot 10^{-7} M_\odot$
yr$^{-1}$, which translates in a gas mass-loss rate of $7 (\pm 1)
\cdot 10^{-5} M_\odot$ yr$^{-1}$, assuming a dust/gas ratio 0.01.
Together with $R_{\mathrm{out}}$ this parameter determines the total
dust mass in the shell surrounding \object{OH 127.8+0.0} and sets it
to be $1.4 \cdot 10^{-3}$ $M_\odot$.

By scaling the SED to the observations, we find that the distance
toward \object{OH 127.8+0.0} is 1800 ($\pm$ 50) pc, which assumes the
stellar luminosity is 6300 $L_\odot$.  This is significantly smaller
than results using the phase lag of the 1612 MHz OH maser in
combination with imaging.  \citet{HH_85_shellsizes} find a distance of
6.98 kpc, and \citet{BJ_90_VLAobs} report 6.21 $\pm$ 1.0 kpc, using
the phase lag determination of \citet{HH_85_shellsizes}. Such a large
distance consequently leads to a very high luminosity ($2.6 \cdot
10^5$ $L_\odot$), more typical for red supergiants than OH/IR stars.
Other studies result in distances of 5.6 kpc \citep{ESW_86_OHIR} and
2.8 kpc \citep{HHF_90_massive}.

\subsubsection{Dust properties}
\label{sec:nonspher}

We find that the dust shell consists of the following dust components:
amorphous olivine (80\% by mass), forsterite (3\%), enstatite (3\%),
metallic iron (4\%) and crystalline water ice (10\%,
$T_{\mathrm{cond}}=150$ K), based on the assumption that all dust
species are present in separate grain populations. The crystalline
silicates forsterite and enstatite are identified in the spectra of
AGB stars \citep{SKB_99_ohir}, and their relative mass fraction is
assumed to be 3\% each, a typical value for AGB stars
\citep{KWD_01_xsilvsmdot}. We did not attempt to fit the actual mass
fraction taken by enstatite and forsterite, which is subject to
further study (Kemper et al., in prep.). The overall shape of the SED
is not affected by assuming a degree of crystallinity of 3\%, since
aside from the sharp resonances in the MIR, the opacities of amorphous
and crystalline silicates are comparable.  All dust is found to be
present in non-spherical (CDE) grains.  The adopted optical constants
of crystalline and amorphous silicates are provided by
\citet{DBH_95_glasses}, \citet{JMD_98_crystalline} and Koike
(priv.comm.).  For crystalline water ice we used the optical constants
derived by \citet{BLW_69_ice} and \citet{W_84_ice}; the optical
constants of metallic iron are taken from \citet{OBA_88_Fe}.

\begin{figure*}
  \resizebox{\hsize}{!}{\includegraphics{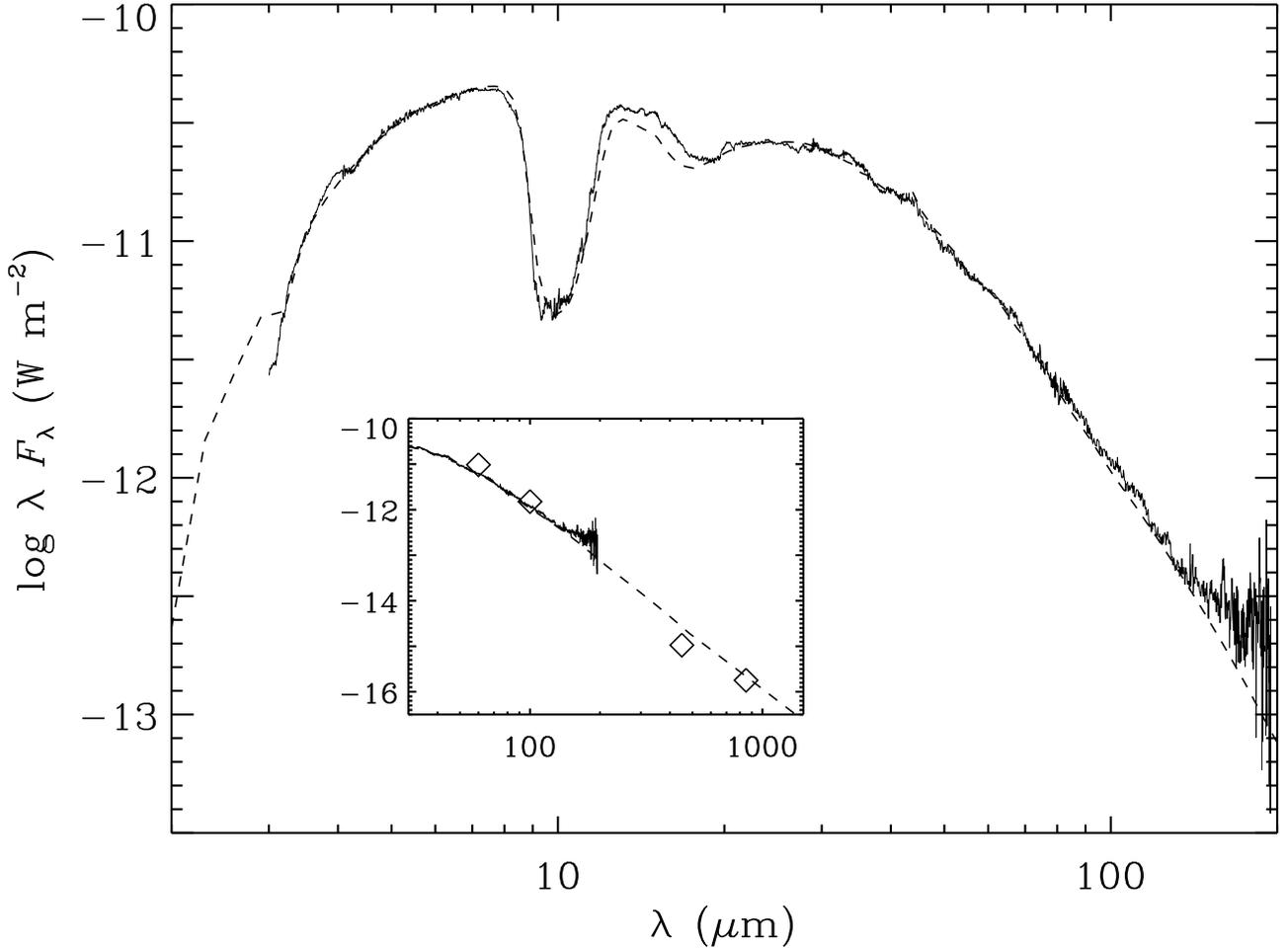}}
\caption{The large panel shows the fit (dashed line) to the ISO spectrum of 
  \object{OH 127.8+0.0} (solid line). In the inset, the same fit is
  presented for the 30 -- 1500 $\mu$m wavelength range.  The diamonds
  indicate the IRAS 60 and 100 $\mu$m and JCMT SCUBA 450 and 850
  $\mu$m photometry points. The error bars on the measurements are
  smaller than the size of the symbols.  A mass fraction of 4\% of
  metallic iron was included in the dust.  See Sect.~\ref{sec:oh127}
  for a description of the model parameters and the dust composition.}
\label{fig:fitoh127}
\end{figure*}

Fig.~\ref{fig:fitoh127} shows the fit to \object{OH 127.8+0.0}.  The
ISO spectroscopy is taken from \citet{SKB_99_ohir} and the 60 and 100
$\mu$m photometry originates from the IRAS Point Source Catalog.
Additional observations with the James Clerk Maxwell Telescope (JCMT)
have been performed to obtain the 450 and 850 $\mu$m photometry points
(Kemper et al., in prep.).  It is clear that if 4\% of the total dust
mass is contained in non-spherical metallic iron grains, the NIR
problem is solved. This is consistent with the average interstellar Fe
and Si abundances \citep{SW_96_depletion} that allow a mass fraction
of 15\% in the form of metallic iron, with respect to amorphous
olivine.

In order to fit the FIR part of the SED, it is required to include a
significant fraction of water ice in the dust shell (10\%) for which a
condensation temperature of 150 K is assumed. This is consistent with
the mass fraction contained in water ice in the oxygen-rich dust shell
of post-AGB star \object{HD 161796} (Hoogzaad et al., in prep.).

\subsection{Comparison with astronomical silicate}
\label{sec:astrosil}

\begin{figure}
  \resizebox{\hsize}{!}{\includegraphics{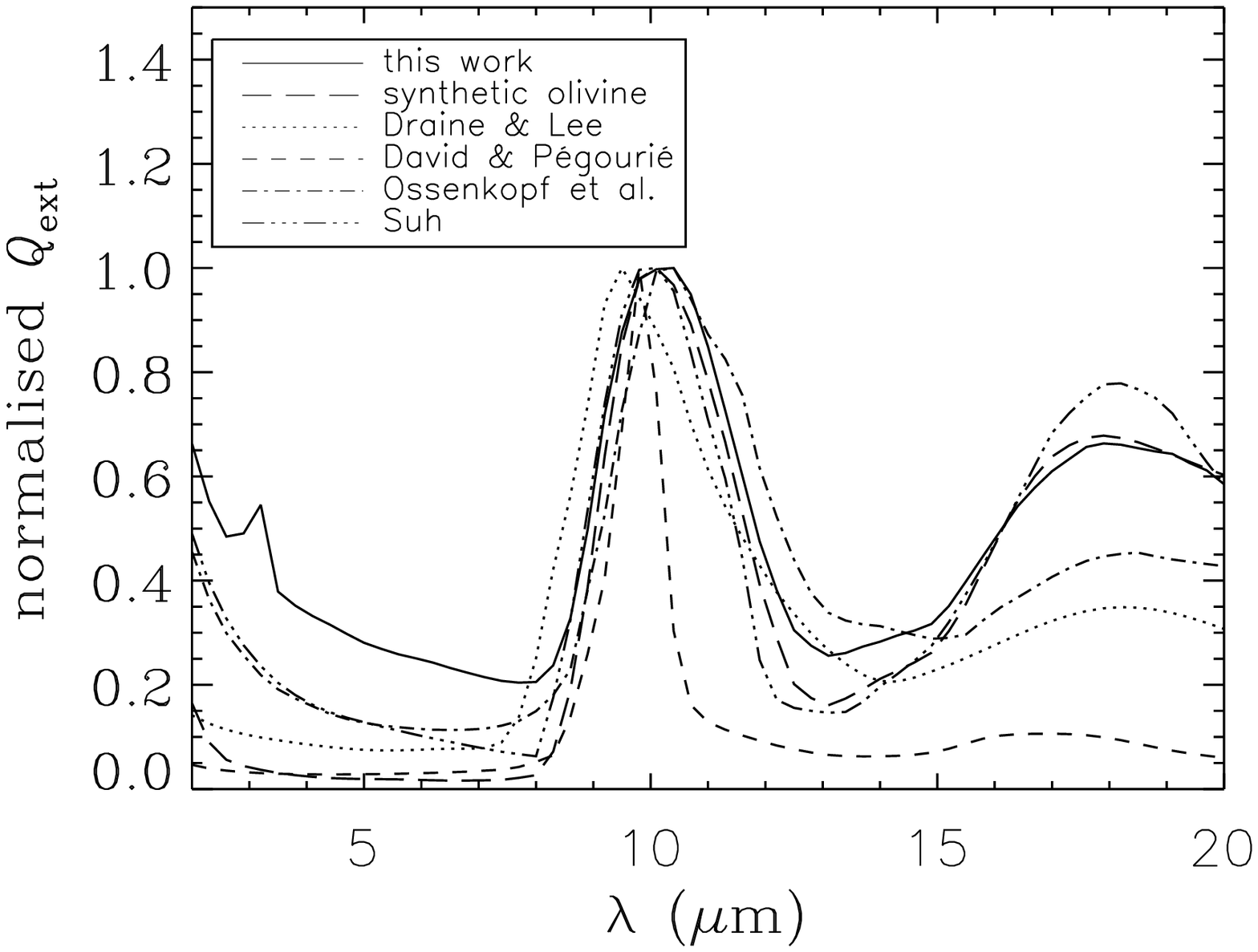}}
\caption{Comparison of the extinction efficiency $Q_{\mathrm{ext}}$ for 
  the composition found for the circumstellar dust shell of \object{OH
    127.8+0.0} (this work), compared with the $Q_{\mathrm{ext}}$ of
  synthesized amorphous olivine \citep{DBH_95_glasses}, and the
  $Q_{\mathrm{ext}}$ for several different astronomical silicates
  \citep{DL_84_optprop,OHM_92_cosmicsilicates,DP_95_cssil,S_99_optprop}.
  All extinction efficiencies are normalised on the 10 $\mu$m peak
  strength to allow comparison of the NIR extinction.}
\label{fig:qcomp}
\end{figure}

The derived total dust extinction toward \object{OH 127.8+0.0} can be
compared with astronomical and laboratory measurements. The extinction
efficiency $Q_{\mathrm{ext}}$ of the individual dust species assuming
CDE have been added proportionately and the resulting total
$Q_{\mathrm{ext}}$ is plotted in Fig.~\ref{fig:qcomp}.  Note that all
curves are normalised to the strength of the 10 $\mu$m feature to
emphasize the differences in the NIR region. To provide some
quantitative insight, Tab.~\ref{tab:astrocomp} presents the extinction
efficiencies at 7 $\mu$m and at 18 $\mu$m with respect to the
extinction efficiencies at 10 $\mu$m.  The long-dashed line shows the
extinction efficiency for a synthesized pure amorphous olivine
\citep{DBH_95_glasses}.  One can see easily that for this species the
NIR extinction is very small compared to the efficiency at 10 $\mu$m.
Also shown in the figure are astronomical silicates derived from ISM
lines-of-sight \citep{DL_84_optprop,OHM_92_cosmicsilicates} and toward
late-type stars
\citep{OHM_92_cosmicsilicates,DP_95_cssil,S_99_optprop}. Only the work
of \citet{OHM_92_cosmicsilicates} assumes non-spherical grains and
uses CDE as a representation of grain shapes. The other efficiencies
are calculated using Mie scattering for spherical particles.

\begin{flushleft}
\begin{table*}
\caption[]{Relative extinction efficiencies at 7 and 18 $\mu$m compared to 
the extinction efficiency at 10 $\mu$m, in addition to Fig.~\ref{fig:qcomp}. 
}
\begin{tabular}{l c c l}
\hline
\hline
optical constants &     $\displaystyle \frac{Q_{\mathrm{ext}}(7\, \mu\mathrm{m})}{Q_{\mathrm{ext}}(10\,\mu\mathrm{m})}$      &       $\displaystyle \frac{Q_{\mathrm{ext}}(18\,\mu\mathrm{m})}{Q_{\mathrm{ext}}(10\,\mu\mathrm{m})}$ & ref.\\
\hline
\object{OH 127.8+0.0} &  0.22     &   0.67    \\
\\
O-rich silicate        &  0.15     &   0.61  & \citet{OHM_92_cosmicsilicates} \\
O-deficient silicate   &  0.12     &   0.47  & \citet{OHM_92_cosmicsilicates} \\
astronomical silicate  &  0.087    &   0.39  & \citet{DL_84_optprop}  \\
warm silicate          &  0.085    &   0.37  & \citet{S_99_optprop} \\
cold silicate          &  0.080    &   0.78  & \citet{S_99_optprop}  \\
astronomical silicate  &  0.041    &   0.12  & \citet{DP_95_cssil}  \\
amorphous olivine      &  0.017    &   0.68  & \citet{DBH_95_glasses} \\
\hline
\hline
\end{tabular}
\label{tab:astrocomp}
\end{table*}
\end{flushleft}

The optical constants derived by \citet{DP_95_cssil} from G and M
supergiants, closely resemble the synthetic olivine in the NIR region.
The 10 $\mu$m resonance is much narrower, while the efficiency at
longer wavelengths is again significantly less compared to our result.
All in all, the David \& P\'{e}gouri\'{e} result gives the poorest
match to our findings.  \citet{S_99_optprop} has derived the optical
constants from AGB stars and splits them into two different types:
\emph{warm} and \emph{cool} silicates, representative for low and high
mass-loss rate AGB stars respectively. This dichotomy assumes a
difference in composition of the bulk material.  In
Fig.~\ref{fig:qcomp}, we present the efficiency for cool dust as this
is most appropriatie for the optically thick dust shell modeled in
this work. The Suh result compares reasonably well with our finding,
though the NIR efficiency is less and the 18 $\mu$m efficiency is
somewhat larger.  \citet{OHM_92_cosmicsilicates} distinguish between
interstellar \emph{O-rich} silicates (olivines) and circumstellar
\emph{O-deficient} silicates (pyroxenes), without discriminating
between optically thin and thick dust shells. Shown in
Fig.~\ref{fig:qcomp} are the \emph{O-deficient} silicates which
represent the most meaningful comparison with our model.  In the NIR
the Ossenkopf et al.~result compares best with our efficiencies,
although the still find a lower $Q_{\mathrm{ext}}$. In the 18 $\mu$m
region their efficiencies also fall below our work.

So, the dust composition determined in this paper provides the only
result that produces sufficient NIR flux. One should note, however,
that our results are of course sensitively dependent on the metallic
iron content of the AGB star, and may vary from source to source,
whereas other studies tried to derive mean values for a sample of
stars.

\section{A consistency check}
\label{sec:lmin12}

The SED of \object{OH 127.8+0.0} is typical for high mass-loss rate
AGB stars, which justifies the calculation of a series of models for
varying mass-loss rates, while the dust composition and other physical
parameters were kept the same. \citet{GS_99_condensation} show that
metallic iron and silicates condense simultaneously in the stellar
outflow, suggesting that the relative mass fraction of iron compared
to the silicates remains constant.  We allowed the inner radius to
change in order to satisfy the constraint that the condensation
temperature $T_{\mathrm{cond}} = 900$ K.

Using the SEDs predicted by the model, the infrared colours in
different stages of the mass-loss evolution on the AGB can be
determined. For example, the relation between the $K-[12$ $\mu$m$]$
colour or $K-L$ colour and the optical depth is a useful diagnostic
tool to determine the mass-loss rate of an AGB star
\citep{LL_96_SED,LW_98_IRMdot}.  We searched for the presence of a
relation between the $L-[12$ $\mu$m$]$ colour index and the mass-loss
rate, since the $L-[12$ $\mu$m$]$ colour can be determined from ISO
SWS spectroscopy directly.  For the L-band (3.6 $\mu$m) we use the
transmission profile provided by \citet{BB_88_JHKLMphotometry}; for
the magnitude calibration an absolute flux density for zero magnitude
of 277 Jy was used \citep{K_83_photometry}.  The standard IRAS
transmission profile and calibration are used to determine the
magnitude in the 12 $\mu$m band. The calculated absolute magnitudes
and the colour indices are summarized in Table~\ref{tab:magnitudes}.

\begin{flushleft}
\begin{table}
\caption[]{Absolute magnitudes and infrared colour index as a function of mass-loss rate, for modelled SEDs with 4\% of the dust mass contained in 
non-spherical metallic iron grains.}
\begin{tabular}{c c c c}
\hline
\hline
$\dot{M}$ ($M_{\odot}$ yr$^{-1}$) & $M_{[12]}$ & $M_L$ & $L-[12]$\\
\hline
$4(-8)$ &$-8.8$ &$-7.4$ &$1.4$\\
$7(-8)$ &$-9.0$ &$-7.4$ &$1.6$\\
$1(-7)$ &$-9.1$ &$-7.4$ &$1.7$\\
$2(-7)$ &$-9.4$ &$-7.5$ &$2.0$\\
$4(-7)$ &$-9.9$ &$-7.5$ &$2.4$\\
$7(-7)$ &$-10.3$        &$-7.5$ &$2.8$\\
$1(-6)$ &$-10.6$        &$-7.6$ &$3.0$\\
$2(-6)$ &$-11.0$        &$-7.7$ &$3.3$\\
$4(-6)$ &$-11.5$        &$-7.9$ &$3.6$\\
$7(-6)$ &$-11.6$        &$-7.9$ &$3.7$\\
$1(-5)$ &$-11.7$        &$-8.0$ &$3.7$\\
$2(-5)$ &$-11.9$        &$-7.8$ &$4.1$\\
$4(-5)$ &$-11.9$        &$-7.4$ &$4.4$\\
$7(-5)$ &$-11.9$        &$-6.8$ &$5.1$\\
$1(-4)$ &$-11.8$        &$-6.3$ &$5.5$\\
$2(-4)$ &$-11.6$        &$-4.9$ &$6.7$\\
$4(-4)$ &$-11.2$        &$-3.2$ &$8.0$\\
$7(-4)$ &$-10.7$        &$-1.7$ &$9.0$\\
$1(-3)$ &$-10.1$        &$-0.7$ &$9.4$\\
\hline
\hline
\end{tabular}
\label{tab:magnitudes}
\end{table}
\end{flushleft}

\begin{figure}
  \resizebox{\hsize}{!}{\includegraphics{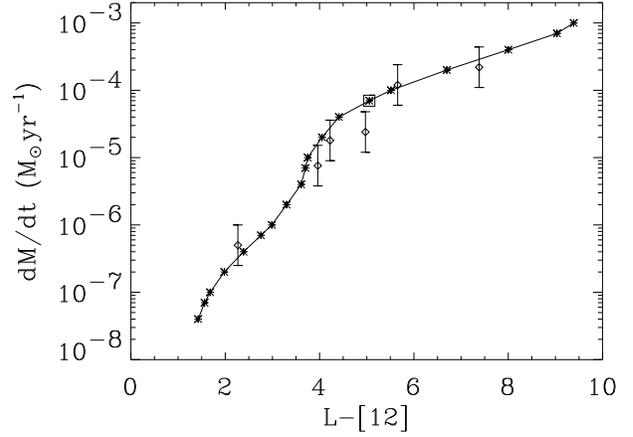}}
\caption{The mass loss rate of oxygen-rich AGB stars as a function
  of the $L-[12$ $\mu$m$]$ colour index, arising from radiative
  transfer calculations, indicated with the asterisks. The fit
  parameters of \object{OH 127.8+0.0} are indicated with a square. The
  diamonds represent the objects from the sample of
  \citet{SKB_99_ohir} with error bars on the mass-loss rate. The
  gas/dust ratio is assumed to be 100. With increasing mass-loss rates
  these are: \object{$o$ Cet}, \object{WX Psc}, \object{CRL 2199},
  \object{OH 104.9+2.4}, \object{OH 26.5+0.6} and \object{OH
    32.8$-$0.3}. The mass-loss rate of \object{$o$ Cet} is taken from
  \citet{LFO_93_CO}.}
\label{fig:l12}
\end{figure}

Fig.~\ref{fig:l12} shows the correlation between the $L-[12$ $\mu$m$]$
colour and the mass-loss rate derived from our model calculations. In
addition, we have included the mass-loss rates and infrared colours of
five stars taken from the sample of \citet{SKB_99_ohir}. The $L-[12$
$\mu$m$]$ colour is obtained directly from the ISO spectroscopy. We
can see that within the accuracy of the mass-loss rate determinations,
a metallic iron fraction of 4\% gives reasonable results, thus
providing a consistency check that adding metallic iron can explain
the infrared colours of other AGB stars as well. From the figure it
becomes also clear that modifications of the iron mass fraction
probably improves the result. To determine the iron mass fraction,
detailed full radiative transfer calculations are necessary.

\section{Discussion}
\label{sec:disc-concl}

The results presented in this work provide the first sound
identification of metallic iron in the circumstellar dust shell around
AGB stars, without violating abundance constraints.
\citet{DDW_00_OHIR} have tried to include metallic iron in their model
calculations of evolved stars, but because they did not consider
non-spherical grains, metallic iron could not be abundant enough to
account for the missing NIR opacity. \citet{HMD_01_VYCMa} have
concluded that metallic iron is necessary to fit the spectrum of
\object{VY CMa}, although all Fe-atoms should be incorporated in
metallic iron, implying that the amorphous silicate is completely
Mg-rich, and even then a large overabundance is required.

From meteoritic and theoretical studies it is known that metallic iron
can condense in a cooling gas of solar composition
\citep[e.g.][]{GL_74_chemicalhistory,LN_79_iron,KH_88_ironbearing}. In
that case, metallic iron forms almost simultaneously with silicates,
since its condensation temperature is only 50 -- 100 K below that of
silicates.  In an oxygen-rich chemistry, metallic iron is stable above
700 K. Below that temperature it will react to form FeS and/or FeO
\citep{J_90_iron}.

An alternative is that it will remain metallic because it is protected
from the oxidizing environment by inclusion in a grain. This can be
achieved through the formation of \emph{iron islands} on the surface
of silicate grains \citep{GS_99_condensation}. During the formation
process of silicates, Fe-atoms in the lattice will be replaced by the
thermodynamically more favourable Mg. The Fe-atoms will migrate to the
surface of the grain to form \emph{iron islands} and stimulate the
condensation of additional Fe-atoms, to continue the growth of the
islands.  \citet{GS_99_condensation} suggest that the islands will
eventually be covered by younger silicate layers, thus creating
silicate grains with platelet shaped, i.e.~non-spherical, metallic
iron inclusions. These inclusions greatly increase the opacity of the
grains in the NIR region.  The model calculations presented in this
work use separate grain populations, rather than metallic iron
inclusions in silicate grains, but still provide an idea of the mass
fraction contained in the iron inclusions.  This result, together with
the inclusion of crystalline water ice and the use of CDE for all
grain shapes provide an important step in disentangling the dust
composition of the circumstellar shell of AGB stars.  No exotic
assumptions were required to match the spectrum. A spherical outflow
of material at constant velocity suffices to produce a good fit to the
SED \emph{if} chemical and shape properties of the grains are treated
in sufficient detail.
 
The presence and formation of metallic iron in astrophysical
environments has been subject to many studies in the past \citep[][and
references herein]{J_90_iron}.  Studies of the interstellar extinction
toward the \object{galactic centre} (GC) revealed a discrepancy in the
3--8 $\mu$m region with the standard dirty silicate opacity
\citep{L_99_gc,LFG_96_gc}. Of course, this is an ISM line-of-sight,
implying that amorphous carbon is an important dust component. On the
other hand, the opacity contribution of amorphous carbon is already
included in the optical constants of the dirty silicate, i.e.~a
graphite-silicate mixture. Thus also in this case, non-spherical
metallic iron grains might explain the missing opacity in the NIR
region.  In addition, it is known that \emph{superparamagnetic}
inclusions, such as metallic iron, in elongated silicate grains can
cause these grains to align when a magnetic field is present
\citep{M_86_alignment}.  The alignment of silicate grains explains the
polarization of starlight passing through the interstellar medium, a
suggestion first made by \citet{ST_51_polarization}.

This theory appears to be supported by studies of interplanetary dust
grains, which are collected by aircraft in the upper atmosphere of the
Earth.  Some of the collected particles are believed to be very
pristine and probably even from interstellar origin, rather than being
reprocessed during the formation of the solar system. These pristine
particles consist of an amorphous matrix material with metal
inclusions, and are referred to as GEMs (glasses with embedded metal
and sulphides).  Metallic iron inclusions are quite common in these
supposedly interstellar grains \citep{B_94_anomalousIDP,M_95_GEMs}.
Whether these grains really originate from the outflows of evolved
stars remains subject to speculation at this moment.
 
\begin{acknowledgements}
  FK, AdK, LBFMW and JB acknowledge financial support from NWO Pionier
  grant 616-78-333. We gratefully acknowledge support from NWO Spinoza
  grant 08-0 to E.P.J.~van den Heuvel.  The James Clerk Maxwell
  Telescope is operated on behalf of the Particle Physics and
  Astronomy Research Council of the United Kingdom, the Netherlands
  Organisation for Scientific Research and the National Research
  Council of Canada.
\end{acknowledgements}

\bibliographystyle{apj} 
\bibliography{ciska}

\end{document}